\documentclass[preprint]{aastex}
\usepackage{graphicx}

\shorttitle{Three new proto-PNe with the 21-$\mu$m feature}
\shortauthors{Cerrigone et al.}

\begin{document}


\title{Identification of three new proto-Planetary Nebulae exhibiting the unidentified feature at 21~$\mu$m}


\author{Luciano Cerrigone\altaffilmark{1}}
\affil{Max-Planck-Institut f\"ur Radioastronomie, Auf dem H\"ugel 69, 53121, Bonn, Germany}
\email{lcerrigone@mpifr.de}

\author{Joseph L. Hora\altaffilmark{2}}
\affil{Harvard-Smithsonian Center for Astrophysics, 60 Garden St, 02138 Cambridge, MA, USA}
\email{jhora@cfa.harvard.edu}

\author{Grazia Umana\altaffilmark{3}, Corrado Trigilio\altaffilmark{3}}
\affil{INAF, Osservatorio Astrofisico di Catania, via S. Sofia 58, 95123 Catania, Italy}
\email{grazia.umana@oact.inaf.it}
\email{corrado.trigilio@oact.inaf.it}

\and

\author{Alexa Hart\altaffilmark{2}, Giovanni Fazio\altaffilmark{2}}
\affil{Harvard-Smithsonian Center for Astrophysics, 60 Garden St, 02138 Cambridge, MA, USA}
\email{ahart@cfa.harvard.edu}
\email{gfazio@cfa.harvard.edu}




\begin{abstract}
Among its great findings, the IRAS mission showed the existence of an unidentified mid-IR feature around 21~$\mu$m.
 Since its discovery, this feature has been detected in all C-rich proto-PNe of intermediate spectral type (A--G) and - weakly - in
a few PNe and AGB stars, but the nature of its carriers remains unknown.
In this paper, we show the detection of this feature in the spectra of three new stars transiting from the AGB to the PN stage 
obtained with the Spitzer Space Telescope. 
Following a recent suggestion, we try to model the SEDs of our targets with amorphous carbon and FeO, which might be responsible for
 the unidentified feature. The fit thus obtained is not completely satisfactory, since the shape of the feature is not well matched. 
%
In the attempt to relate the unidentified feature to other dust features, we retrieved mid-IR spectra of all the 21-$\mu$m sources currently known from ISO and Spitzer on-line archives and noticed a correlation between the flux emitted in the 21-$\mu$m feature
 and that emitted at 7 and 11 $\mu$m (PAH bands and HAC broad emission). Such a correlation may point to a common nature of the carriers.
\end{abstract}



\section{Introduction}
One of the major results of the \textit{Infrared Astronomical Satellite} (IRAS) mission came from its Low Resolution Spectrometer (LRS) with the detection of a new broad feature around 21~$\mu$m \citep{kwok89} in the spectra of stars that have gone through the Asymptotic Giant Branch (AGB) phase. The existence of this feature was not recognized at once, because it is weak in the first object in which it was observed, IRAS~22272+5435, which was listed in the LRS Atlas as having a low-temperature continuum and silicate absorption \citep{volk99}. In the first paper about this feature, four sources are listed (namely IRAS 07134+1005, 23304+6147, 22272+5435, and 04296+3429); subsequent ground-based (United Kingdom InfraRed Telescope), airborne (Kuiper Airborne Observatory), and space-based (Infrared Space Observatory) observations confirmed those detections and also added new members to the group of 21-$\mu$m emitters.  Observations  with the \textit{Infrared Space Observatory} (ISO) led to conclude that the feature actually peaks around 20.1~$\mu$m \citep{volk99}, although it is still today commonly referred to as the \lq\lq 21~$\mu$m feature\rq\rq. 

The feature has been detected almost exclusively in C-rich proto-Planetary Nebulae (PPNe), interme\-diate-mass stars (1--8 M$_\odot$) transiting from the AGB to the Planetary Nebula (PN) stage. 

During the AGB phase, stars lose most of their initial mass, with mass loss 
rates ranging from 10$^{-7}$ up to 10$^{-4}$~M$_\odot$~yr$^{-1}$. The AGB stellar atmospheres are subject to periodic He-flashes, which imply
that enriched material produced by nucleosynthesis in the bottom layers is carried to the surface. If this dredge-up of the star is 
efficient enough, the chemistry of the cooled  material in the circumstellar envelope (CSE) will be based on C rather than O, as normally expected.
Depending on their $[$C$]/[$O$]$ ratio, we can then distinguish between O-rich and C-rich CSEs.

 When the star has evolved past the tip
 of the AGB, the intense mass loss that characterizes the AGB has ceased, but the dust and gas in the CSE is typically 
so thick that the star cannot be detected at optical 
wavelengths. The expansion and subsequent dilution of the envelope 
leads to the optical detection of a post-AGB star, which typically has a double-peaked Spectral Energy Distribution (SED),
 as the result of the radiation from the central source and its CSE. 

The central star then evolves towards hotter temperatures,
while its CSE continues to expand and dilute. If the central star does not evolve too slowly, a few thousand 
years after the AGB phase it is hot enough to ionize its CSE (T$_{\star}$ $\sim$20--30$\times 10^3$ K), which can lead to the destruction of molecules and even dust grains.
The point at which the ionization of the CSE occurs is in fact considered as the beginning of the Planetary Nebula phase.

All of the objects showing the 21~$\mu$m feature are C-rich and metal poor with enhanced abundances of s-processed elements 
\citep{hrivnak09}, and typically with F--G spectral type. Besides the unidentified feature, their mid-IR spectra typically show  
a broad feature around 30~$\mu$m that is usually attributed to MgS \citep{molster} and has been found to consist of two features centered at 26 and 33 $\mu$m, 
though such a distinction is not observed in all sources \citep{hrivnak09}.

As of this writing, 16 PPNe are known to exhibit the unidentified 21-$\mu$m feature \citep{hrivnak09}. 
Besides these sources, the feature has been detected weakly in three PNe \citep{hony01,volk02} and three AGB stars \citep{volk00}. 
It has not been observed in IRAS 01005+7910 \citep{cerrigone09}, a C-rich PPN with a hot central star (B0 I) that has already 
started to ionize its circumstellar shell \citep{cerrigone10}. Other hot post-AGB stars showing features of  Polycyclic Aromatic Hydrocarbons (PAH) in their mid-IR spectra 
and therefore likely to be C-rich do not exhibit the feature either \citep{cerrigone09}. This leads to the conclusion that the carrier must 
form during the AGB phase or very early afterwards, and is probably easily destroyed as the radiation field of the central star hardens.

Since its first detection, many different molecules have been considered as possible carriers of the 21-$\mu$m feature. 
The connection with the C-rich nature of the envelopes makes it obvious to take into account C-bearing molecules, for instance TiC, 
doped-SiC, PAH, Hydrogenated Amorphous Carbon (HAC), and fullerenes. Yet, non C-bearing molecules 
have been indicated as possible carriers as well, for example SiS$_2$, FeO, Fe$_2$O$_3$, and Fe$_3$O$_4$  \citep{zhang}. The main issues 
that arise with almost all of the proposed carriers are the presence of secondary features that are not detected in the astrophysical spectra, or the anomalous abundances necessary to account for the observed intensity of the feature.

In this paper, we present observations obtained within a program aimed at characterizing the dust
emission in objects transiting from the AGB to the PN stage with A--G spectral types, as a follow-up to our project presented in \citet{cerrigone09}, which focused on  sources with hot central stars (B spectral type). While the whole sample of transition targets will be presented in a separate paper along with data from other Spitzer programs (Hart et al, in preparation), here we will focus on three new sources that show the unidentified 21-$\mu$m feature. These are the only targets in our sample, where we find evidence for this feature.

\section{Observations}
The new observations presented in this paper were carried out with the Infrared Array Camera \citep[IRAC;][]{irac} and the 
InfraRed Spectrograph \citep[IRS;][]{irs} on-board the Spitzer Space Telescope \citep{spitzer}, within Program 50116 (PI: G. Fazio)
 in March and April 2009. 
The IRS had four modules with different spectral resolutions. The Short-Low (SL) and Long-Low (LL) modules covered the 5.2--14.5 and 14.0--38.0 $\mu$m ranges respectively, with spectral resolution R$\sim$60--120. The Short-High (SH) and Long-High (LH) modules covered the 9.9--19.6 and 18.7--37.2 $\mu$m ranges with R$\sim$600. Our spectral data were typically acquired in low resolution mode. 
Targets with IRAS fluxes that would imply saturation with the IRS at longer wavelengths were observed in low resolution only between 5
 and 14~$\mu$m (SL modules), while the 10--38~$\mu$m range was observed in high spectral resolution (SH and LH modules). 
In the latter case, a dedicated background observation was also performed for each target. For the three sources presented here, 
low resolution observations were performed for IRAS 13245-5036 and 15482-5741, while low and high resolution data were collected for 
IRAS 14429-4539. IRAS 13245-5036 was not centered on the SL slit, therefore only LL data are available for this target.

Basic Calibrated Data were retrieved from the Spitzer web archive. The observations of IRAS 13245-5036 were processed with the 
18.7.0 pipeline by the Spitzer Science Center (SSC), while version 18.18.0 was applied to 14429-4539 and 15482-5741. 
The data were reduced following the standard recipe in the on-line IRS Cookbook: 
the background subtraction was performed from the off-target observations for low resolution data (i.e., when one module is on the target, 
the other one is observing off-source) and from the dedicated background observation for the high resolution data sets.
 The spectra were cleaned using IRSCLEAN, extracted and averaged together in SMART\footnote{SMART was 
developed by the IRS Team at Cornell University and is available through 
the Spitzer Science Center at Caltech.} \citep{smart}. The high spectral resolution data were extracted over the full slit, while
the optimal extraction algorithm \citep{smart2} was applied to the SL and LL observations.


The IRAC observations were performed in high dynamic range mode with a small dither pattern, as very extended emission 
was not expected. In this observing mode, two sets of BCDs are created by default, corresponding to a 
long and a short exposure; in our case the exposures were respectively 12s and 0.6s long. All of the IRAC BCDs retrieved from 
the Spitzer archive were processed with the S18.18.0 pipeline. The BCDs were treated for artifact correction with the IDL code 
available from the SSC web site and then multi-frame aperture photometry was performed with the MOsaicker and Point 
source EXtractor (MOPEX) using a 10-pixel aperture and 12-20 pixel annulus, which is the standard for the IRAC calibration. 
As the IRAC absolute flux calibrators are known to about 2--3\%, and IRAC repeatability is at about the 1.5\% level, we assume  in the 
following 5\% as our photometric error. Since our targets are quite bright for IRAC, the photometry extraction was performed on the 
short frames, to avoid any saturation effect, which was evident in IRAS 14429-4539 at 8.0~$\mu$m even in the short frames. 
The flux densities are listed in Table~\ref{tab:irac}. 
\begin{table}
\centering
\begin{tabular}{lcccc}
\hline\hline\noalign{\smallskip}
Target      &  3.6~$\mu$m &  4.5~$\mu$m & 5.8~$\mu$m  & 8.0~$\mu$m  \\
      &  mJy &  mJy & mJy & mJy \\
\hline\noalign{\smallskip}
13245-5036  & 21.08 & 17.24 & 71.21 & 440.9 \\
14429-4539  & 573.3 & 1132 & 3256 & -- \\
15482-5741  &  40.41 & 32.37 & 47.39 & 209.3 \\
\hline
\end{tabular}
\caption{Photometry of our targets obtained with IRAC. The absolute accuracy is about 5\%.}
\label{tab:irac}
\end{table}

\section{Spectra and SED modeling}
\begin{figure*}
\epsscale{0.8}
\plotone{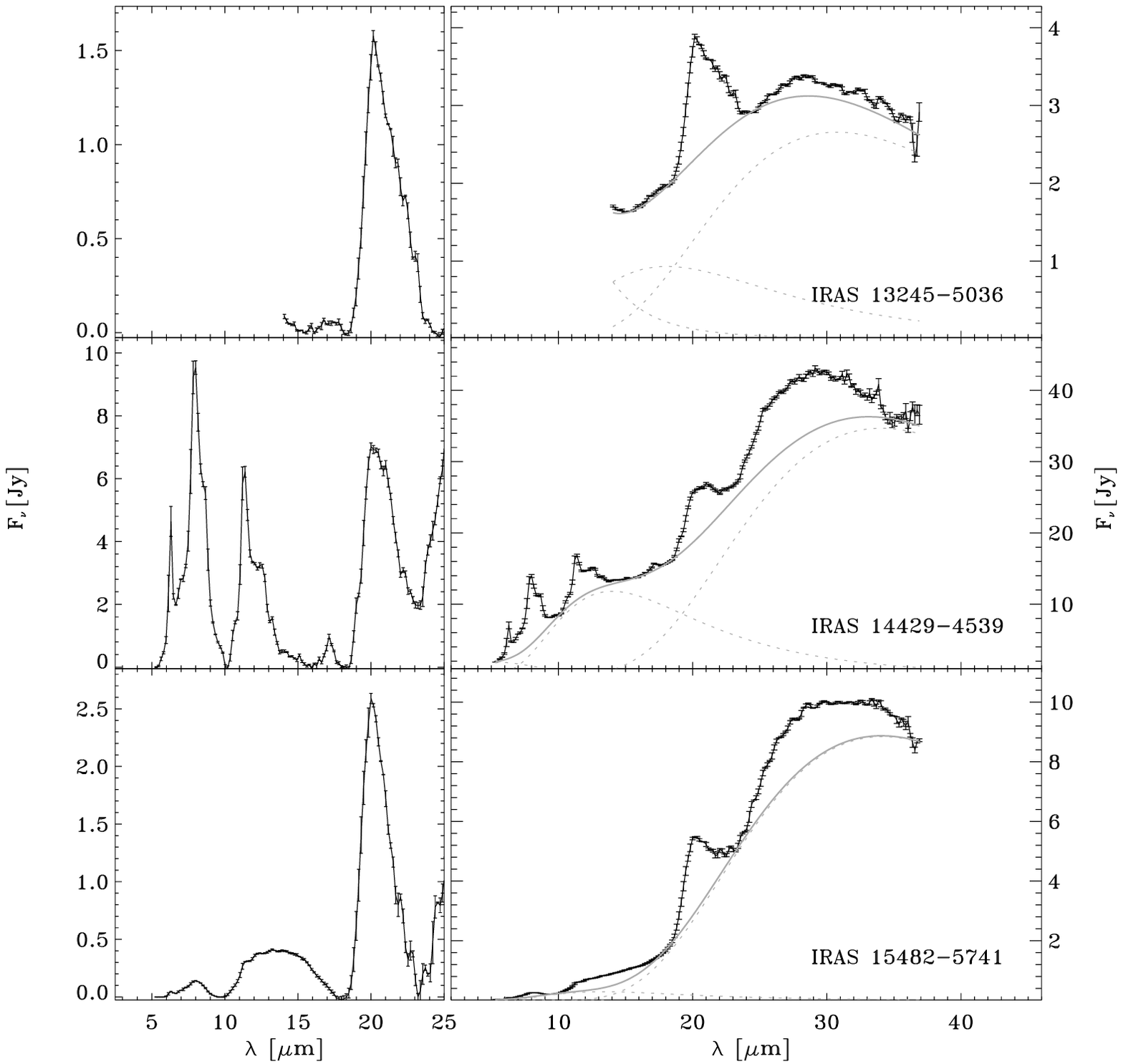}
\caption{The spectra obtained with Spitzer are shown on the \textit{right} side, where a gray-body fit to the continuum (light-gray solid line) and the single gray bodies (light-gray dashed lines) are included. On the \textit{left} side, a close-up of the 5--25~$\mu$m range, after subtraction of the continuum.}
\label{fig:spectra}
\end{figure*}

\begin{figure}
\epsscale{0.6}
\plotone{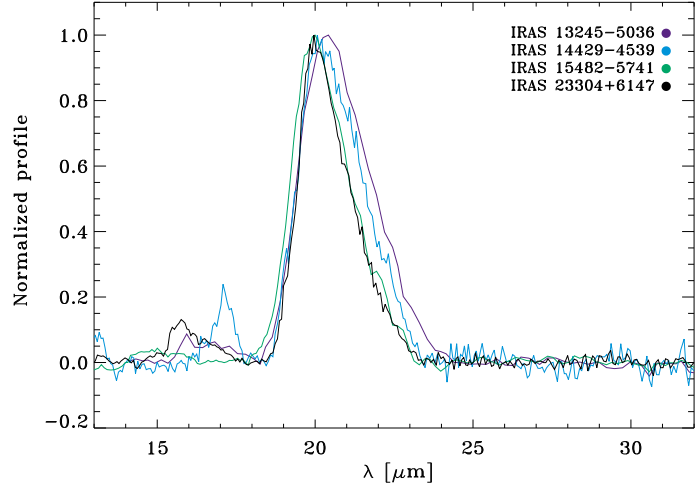}
\caption{Normalized profiles of the 21-$\mu$m feature in our targets compared with that in IRAS 23304+6147.}
\label{fig:solo21}
\end{figure}

Figure~\ref{fig:spectra} shows the results of our spectral observations. For IRAS 13245-5036, only long-wavelength data are shown, 
because the short-wavelength slit (perpendicular to the long-wavelength one) was off the target. The 21-$\mu$m feature is evident 
in all of the three targets. IRAS~14429-4539 and 15482-5741 also clearly show the mid-IR features attributed to PAHs, although the 
balance among the features in the PAH bands seems to be very different in the two sources. In IRAS 14429-4539 these features resemble
 closely those detected in PNe and other C-rich stars, while usually this is not the case for the 21-$\mu$m sources, where the PAH features in the 10--17-$\mu$m range (due to CH out-of-plane bending modes) are often blended together in a plateau (see for example IRAS 22223+4327 in Figure~\ref{fig:old2}) and are as strong or even stronger than the features in the 5--10-$\mu$m range (due to CC stretch and CH in-plane bending modes; \citet{tielens}), as for example in IRAS 22574+6609.
Figure~\ref{fig:solo21} shows the normalized profiles of the 21-$\mu$m features in our targets and in IRAS~23304+6147, which can
 be considered as prototypical. As usually observed, the feature has a relatively regular spectral shape, with no hints for substructure.

To estimate the relative contributions of the PAH features within the emission bands, we have fitted the spectra using the 
publicly-available IDL tool PAHfit
\citep{pahfit}, specifically created to fit PAH features in low-resolution Spitzer spectra. The spectra were binned to a step size of 
0.169~$\mu$m. 

\begin{figure*}
\centering
\includegraphics[width=0.49\textwidth]{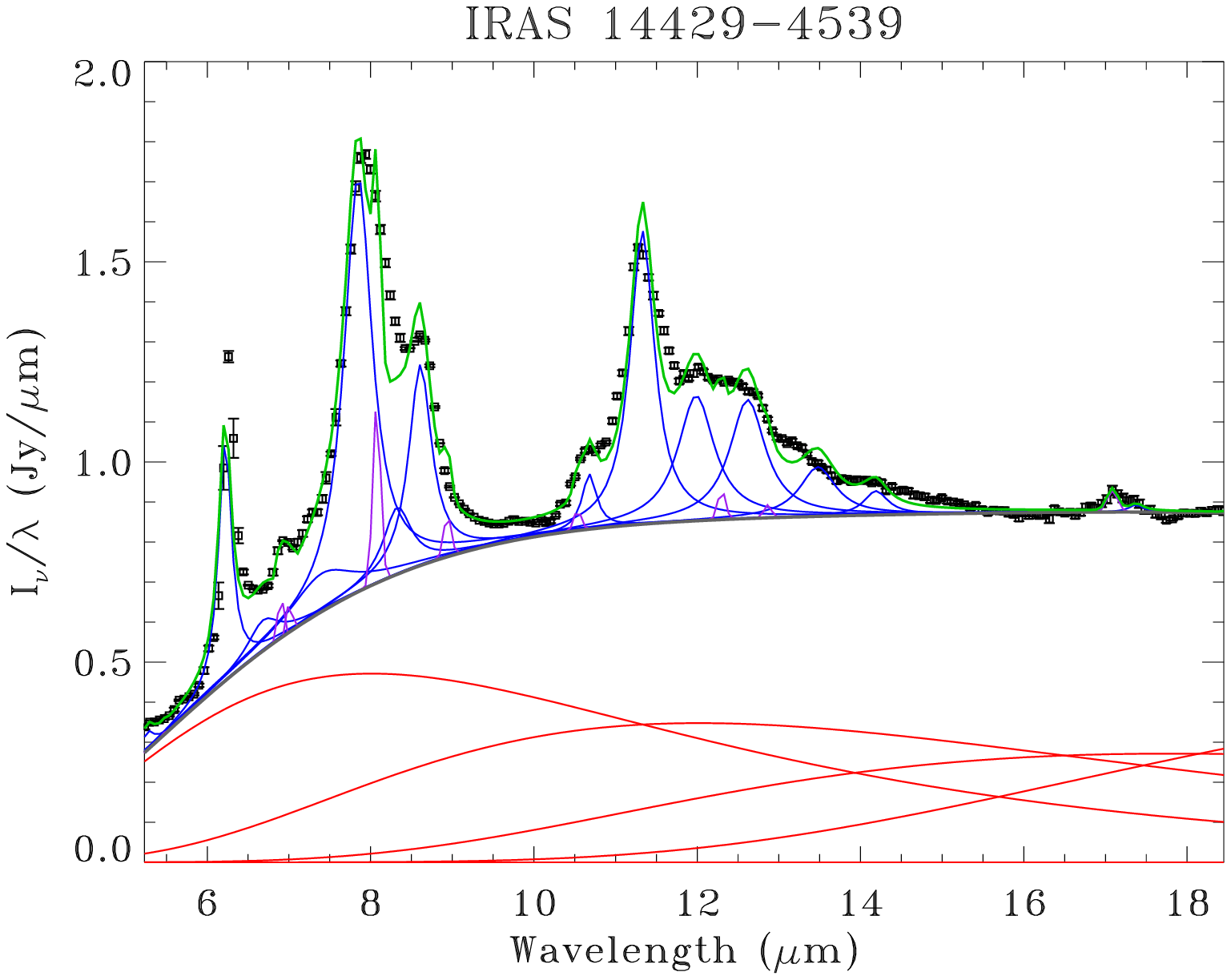}  \includegraphics[width=0.49\textwidth]{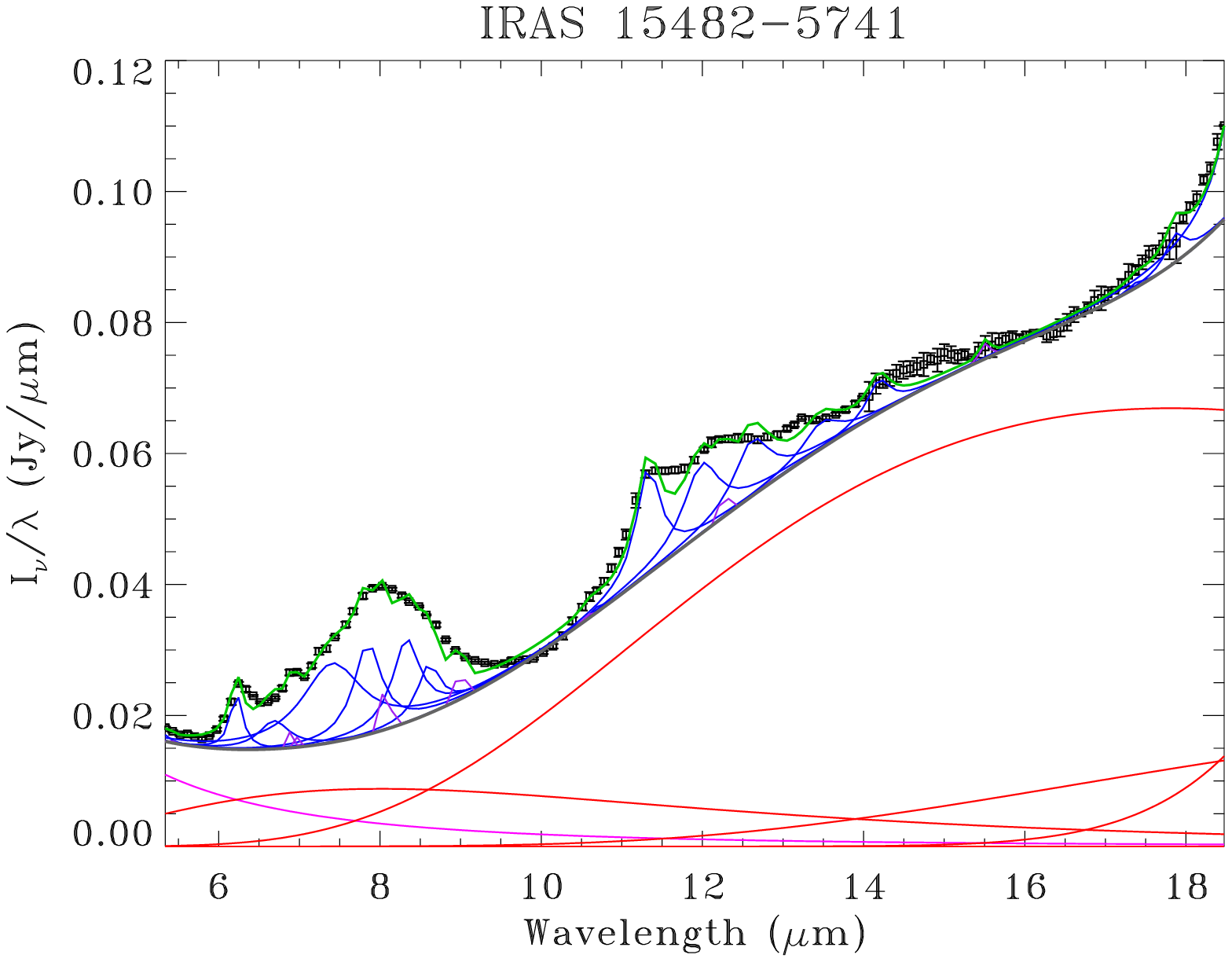}
\caption{Fits to the PAH features obtained with the IDL tool PAHfit. Single gray bodies are shown in red and their total as a thick gray line. PAH features are shown in blue and atomic lines in purple. The sum of all components is shown as a green line.}
\label{fig:pahfit}
\end{figure*}

PAHfit performs a fit to the spectra assuming the continuum is given by a combination of gray bodies with set temperatures of 300,
200, 135, 90, 65, 50, 40, and 35 K and a stellar continuum given by a Planck curve at a temperature input by the user. 
The method of fitting gray bodies is based on the fact that at these wavelengths the underlying continuum is a combination of radiation from the central star and the dust surrounding it. The dust layer extends outwards as far as it cannot be distinguished from the ISM, therefore it is characterized by a continuum of temperatures, typically in the range of a few 100 K to 10--20 K.  
In our two cases,
the software used gray bodies at 300, 200, and 135 K for IRAS 14429-4539 and at 300, 135, 90, and 35 K for 15482-5741. The fact that PAHfit returns different temperature combinations for the two sources should not be over-interpreted, as we are still fitting only a small spectral range.
For the central stars, since all of our targets have been optically classified in \citet{suarez}, we adopt the temperatures 
approximately corresponding to their spectral classifications: 7000 K for 14429-4539 and 6500 K for 15482-5741.
 We list the output obtained from PAHfit in Tables \ref{tab:pahfit1} and \ref{tab:pahfit2}  and the fits are shown in Figure~\ref{fig:pahfit}.
  Besides PAH features, PAHfit also tries to fit several atomic lines, which we have included in our tables. These lines are blended with the PAH features: their presence is supposed to obtain a good fit. In IRAS 15482-5741, two atomic lines are particularly weak and marginal: no error estimates are given for these lines.
\begin{table}
\centering
\begin{tabular}{lccc}
\hline\hline\noalign{\smallskip}
Feature & Wavelength & Peak intensity & Flux \\
        &  $\mu$m   &  Jy            &  $10^{-15}$~W~m$^{-2}$ \\
\hline\noalign{\smallskip}
\multicolumn{4}{l}{\textit{PAH features}}   \\
   5.3   &   5.27  &  $ 0.24  \pm   0.02 $   & $ 7.4 \pm  0.5 $  \\
   6.2   &   6.22  &  $  3.84 \pm   0.06 $   & $ 87  \pm  1   $   \\
   6.7   &   6.69  &  $ 0.515 \pm  0.009 $   & $ 25.4\pm  0.4 $  \\
   7.4   &   7.42  &  $ 0.71  \pm   0.03 $   & $ 56  \pm  2   $  \\
   7.8   &   7.85  &  $  8.16 \pm   0.06 $   & $ 260 \pm  2   $   \\
   8.3   &   8.33  &  $  1.40 \pm   0.02 $   & $ 39.4\pm  0.5 $   \\
   8.6   &   8.61  &  $  4.32 \pm   0.02 $   & $ 92.2\pm  0.4 $   \\
  10.7   &   10.68 &  $   1.48\pm    0.02$   & $ 13.0\pm   0.2$  \\
  11.3   &   11.33 &  $   8.31\pm    0.07$   & $ 111 \pm   1  $   \\
  12.0   &   11.99 &  $   3.75\pm    0.04$   & $ 66.3\pm   0.6$   \\
  12.6   &   12.62 &  $   3.75\pm    0.04$   & $ 58.7\pm   0.6$   \\
  13.5   &   13.48 &  $   1.67\pm    0.04$   & $ 23.3\pm   0.6$  \\
  14.2   &   14.19 &  $  0.86 \pm    0.07$   & $ 7.1 \pm   0.6$  \\
  17.4   &   17.37 &  $  0.3  \pm     0.1$   & $ 0.9 \pm   0.4$ \\
\multicolumn{4}{l}{\textit{Atomic lines}}   \\
      H$_2$ S(5)     &   6.90   & $   1.09 \pm  0.07$ & $ 4.24 \pm 0.27  $  \\
      $[$\ion{Ar}{2}$]$  &   7.01   & $  0.73  \pm  0.04$ & $ 2.77 \pm 0.17  $   \\
      H$_2$ S(4)     &   8.10   & $   3.66 \pm  0.09$ & $19.8 \pm0.5   $  \\
      $[$\ion{Ar}{3}$]$ &   8.94   & $  0.93  \pm  0.02$ & $4.06 \pm0.09   $  \\
      $[$\ion{S}{4}$]$   &  10.53   & $  0.54  \pm  0.03$ & $1.71 \pm0.10   $   \\
      H$_2$ S(2)     &  12.30   & $  0.94  \pm  0.05$ & $2.18 \pm0.12   $   \\
      $[$\ion{Ne}{2}$]$  &  12.86   & $  0.41  \pm  0.06$ & $0.88 \pm0.13   $   \\
      H$_2$ S(1)     &  17.08   & $  0.90  \pm  0.14$ & $1.52 \pm0.24   $  \\
          \hline
\end{tabular}\caption{Output obtained from the IDL tool PAHfit for IRAS 14429-4539 .}\label{tab:pahfit1}
\end{table}

\begin{table}
\centering
\begin{tabular}{lccc}
\hline\hline\noalign{\smallskip}
Feature & Wavelength & Peak intensity & Flux \\
        &  $\mu$m   &  Jy            &  $10^{-15}$~W~m$^{-2}$ \\
\hline\noalign{\smallskip}
\multicolumn{4}{l}{\textit{PAH features}}   \\
    5.3   &     5.27 &  $ 0.008   \pm  0.002 $ &      $ 0.25 \pm   0.07$   \\
    6.2   &     6.22 &  $  0.053  \pm  0.002 $ &      $ 1.21 \pm   0.04$   \\
    6.7   &     6.69 &  $  0.029  \pm  0.001 $ &      $ 1.42 \pm   0.06$   \\
    7.4   &     7.42 &  $  0.090  \pm  0.002 $ &      $ 7.2  \pm   0.1 $   \\
    7.8   &     7.85 &  $   0.110 \pm  0.002 $ &      $ 3.51 \pm   0.07$    \\
    8.3   &     8.33 &  $   0.106 \pm  0.002 $ &      $ 3.00 \pm   0.05$    \\
    8.6   &     8.61 &  $  0.066  \pm  0.001 $ &      $ 1.40 \pm   0.03$   \\
   10.7   &     10.68  & $ 0.009   \pm 0.004  $ &      $ 0.08 \pm   0.04$   \\
   11.3   &     11.33  & $   0.185 \pm 0.005  $ &      $ 2.46 \pm   0.06$       \\
   12.0   &     11.99  & $   0.128 \pm 0.004  $ &      $ 2.27 \pm   0.07$       \\
   12.6   &     12.62  & $   0.110 \pm 0.006  $ &      $ 1.72 \pm   0.09$    \\
   13.5   &     13.48  & $  0.055  \pm 0.004  $ &      $ 0.77 \pm   0.05$     \\
   14.2   &     14.19  & $  0.071  \pm 0.007  $ &      $ 0.59 \pm   0.06$     \\
   17.4   &     17.37  & $  0.01   \pm  0.01  $ &      $0.045 \pm   0.04$   \\
  \multicolumn{4}{l}{\textit{Atomic lines}}   \\
      H$_2$ S(5)  &  6.916 & $  0.0648  \pm n.a $ &  $ 0.22 \pm  n.a. $  \\
       $[$\ion{Ar}{2}$]$  &  7.002 &  $ 0.0336  \pm n.a. $&  $0.11 \pm n.a. $ \\
      H$_2$ S(4)  &  8.076 &  $ 0.073 \pm 0.003  $&  $0.39 \pm 0.02 $ \\
      $[$\ion{Ar}{3}$]$  &  8.994 &  $ 0.060  \pm0.003  $&  $0.26 \pm 0.01 $ \\
      H$_2$ S(2)  & 12.264 &  $ 0.06  \pm  0.01  $&  $0.15 \pm 0.02 $  \\
      $[$\ion{Ne}{3}$]$  & 15.505 &  $ 0.06  \pm  0.02  $&  $0.12 \pm 0.04 $ \\
      H$_2$ S(1)  & 17.085 &  $ 0.011 \pm0.003   $&  $0.018 \pm 0.005 $  \\

\hline
\end{tabular}        
\caption{Output obtained from the IDL tool PAHfit for IRAS 15482-5741.}\label{tab:pahfit2}
\end{table}

\begin{figure*}
\centering
\includegraphics[width=\textwidth]{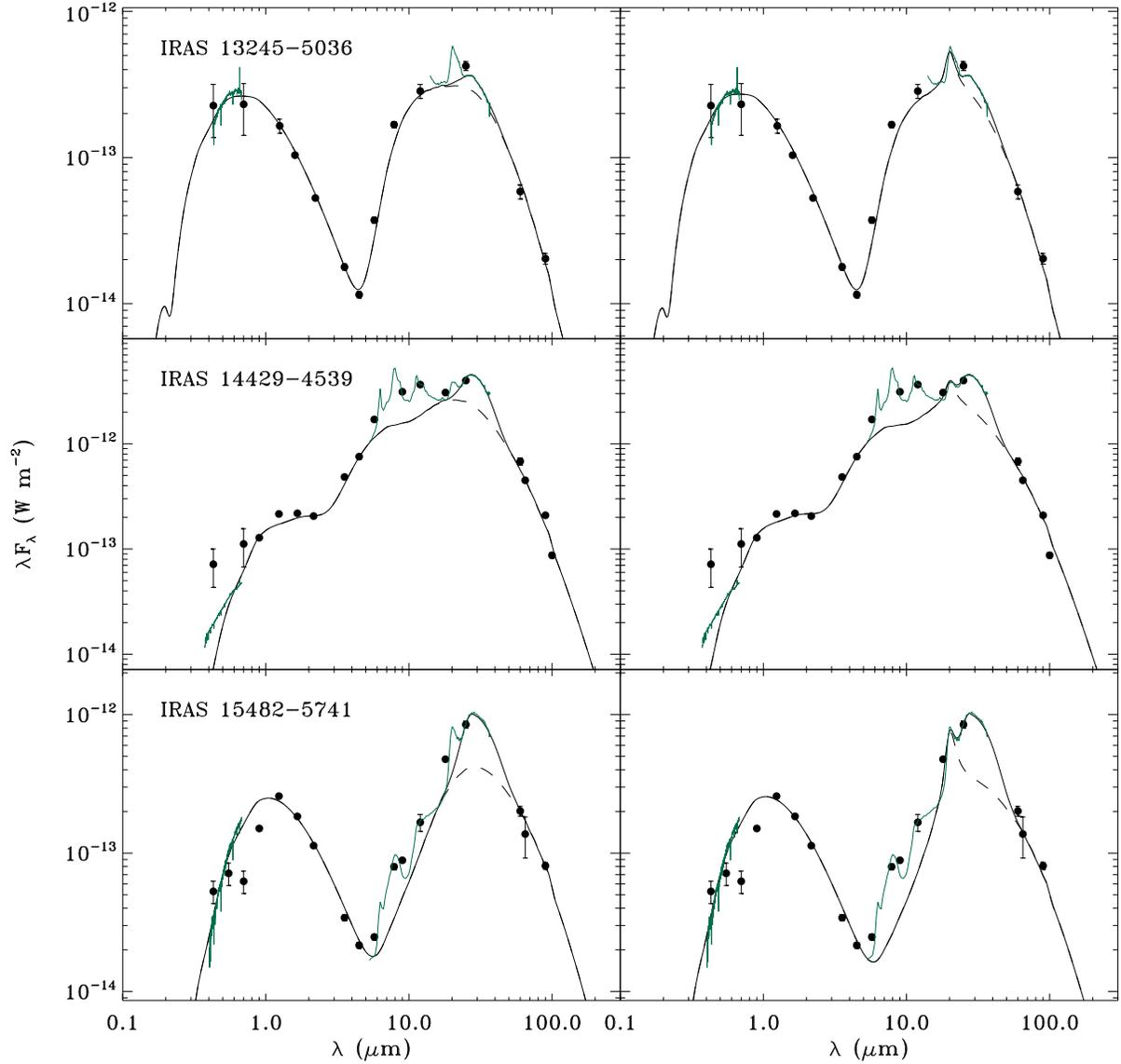}
\caption{DUSTY models obtained for our targets. The DUSTY output is shown as a dashed line; a solid line representing the final model SED (DUSTY output plus an empirical fit to the 30~$\mu$m MgS feature) has been also plotted. On the \textit{left} side, only amorphous carbon grains were used, while the models on the \textit{right} side include a mixture of amorphous C and FeO. Optical spectra from \citet{suarez} and our Spitzer spectra are shown in green.}
\label{fig:dusty}
\end{figure*}

As a further tool to analyze the properties of the envelopes in these stars, we performed radiative transfer modeling using the code DUSTY
 \citep{dusty}, which solves the equation of radiation transfer in a spherical shell of dust grains surrounding a central star. 
 For the grain sizes, we adopted a standard MRN \citep{mrn} distribution of radii varying as $a^{-3.5}$ with a$_\mathrm{min}=0.005$~$\mu$m and 
a$_\mathrm{max}=0.25$~$\mu$m, $a$ being the radius of each grain (assumed spherical). A radial density distribution going as 
$r^{-2}$ was implemented in all models and Planck curves were used to model the radiation from the central stars. The choice of a density distribution varying as r$^{-2}$ is based on the expectation of a constant mass-loss rate. From the equation of mass continuity, we have that
\begin{equation}
 \rho(r)=\frac{\dot{M}}{4\pi r^2 V_\infty}
\end{equation}
where $V_\infty$ is the wind terminal velocity. As a consequence, if the mass loss process is steady, the CSE will be characterized by $\rho(r)\sim r^{-2}$. Towards the end of the AGB, the star is supposed to undergo a short (a few 10$^3$ yr) period of enhanced mass loss called superwind phase, therefore departures from an r$^{-2}$ distribution may be found in these stars. 
Once a satisfactory fit to the data was achieved, we reddened the modelled SED according to \citet{cardelli}, taking A$_\mathrm{V}$ 
as a free parameter. We also summed to the DUSTY model,  an empirical fit to the 30-$\mu$m MgS feature obtained with an asymmetrical Gaussian.
 This was done to better reproduce the emission under the 21-$\mu$m feature, which we tried to model as due to FeO, as recently 
suggested by \citet{zhang}. DUSTY handles mixtures of chemical species as simulating a single-type grain constructed from an average.
 For both amorphous carbon and FeO, we used the default sets of optical constants that come with DUSTY. 
The details of the modeling are summarized in Table~\ref{tab:dusty} and the models are shown in Figure~\ref{fig:dusty}. 

The observational points used in modeling are from our IRAC photometry and the GSC2.2 \citep{gsc2}, NOMAD \citep{nomad}, DENIS \citep{denis}, 2MASS \citep{2mass}, AKARI \citep{akari}, and IRAS \citep{iras} catalogs.

\begin{table}
\centering
\begin{tabular}{lccccccc}
\hline\hline\noalign{\smallskip}
Target & T$_\star$ & [FeO] & $\frac{\mathrm{L}}{\mathrm{d}}$ & T$_\mathrm{dust}^\mathrm{in}$ & $\frac{\mathrm{R}_\mathrm{out}}{\mathrm{R}_\mathrm{in}}$ & $\tau_\mathrm{V}$ & A$_\mathrm{V}$ \\
 & K & \% & L$_\odot$~kpc$^{-1}$ & K &  &  &  \\
\hline\noalign{\smallskip}
13245-5036 & 8500 & 9 & -- & 220 & 1.1 & 0.25 & -- \\
           &   -- & 9 &  267 & 100 & 1.1 & 0.25 & 1.05 \\
14429-4539 & 7000 & 7 & --    & 500 & 10 & 0.6 & -- \\
           & --   & 7 & 673 & 160 & 20 & 1.6 & 4.1 \\
15482-5741 & 6500   & 20 & --   & 255 & 10 & 0.07 & -- \\
           &  --    & 20 & 267  & 120 & 30 & 0.55 & 2.3 \\
\hline
\end{tabular}
\caption{Details of the modeling performed with DUSTY. Except abundances, the same parameter values were used to generate models with and without FeO.}
\label{tab:dusty}
\end{table}

As often observed in post-AGB stars, we found that while the far-IR observational points can be fitted well, the model typically underestimates the radiation at near- and mid-IR wavelengths. This implies that a larger amount of hot dust is present in the CSE than what can be found in a smooth spherical shell. To account for such an emission component, we have  tried different density distributions ($r^{-3}$ and $r^{-1}$), but found that a satisfactory fit can be obtained by running the code twice with $\rho(r) \sim r^{-2}$. In the first run we try to reproduce the data at wavelengths below about 10 $\mu$m, then in the second run we use as a central source the SED obtained as the output of the first run, which contains the central star and the hot dust. Such a density picture resembles a circumstellar environment where the mass-loss process is not constant 
with time, but subject to episodic enhancements. The presence of a  hot-dust component can also be interpreted as an indication for a circumstellar (circumbinary?) disk \citep{deruyter}.

Our attempt to reproduce the 21-$\mu$m feature with FeO following \citet{zhang} is not satisfactory. In particular, the spectral shape 
is not well reproduced and large abundances ($\sim$10\% in mass) of FeO are necessary. Software allowing for a continuous range of 
ellipsoidal grain shapes has been found to account better for spectral shapes of dust features in general \citep{kemper}. It is possible
 that a similar approach may improve the match between the model and the observational data also in our targets. Currently, this is not 
possible with DUSTY.

\section{A link between the 21-$\mu$m feature and hydrocarbon emission?}
It is known that the 21-$\mu$m feature is often - but not always - observed along with mid-IR PAH bands \citep{volk99}. 
Another source of mid-IR emission in the form of a very broad feature resembling an underlying continuum is given by HAC, 
whose molecular chains are considered as the precursors of PAHs.
Given this qualitative observational link between these features, we have searched for a quantitative correlation.

\begin{figure*}
\epsscale{0.63}
\plotone{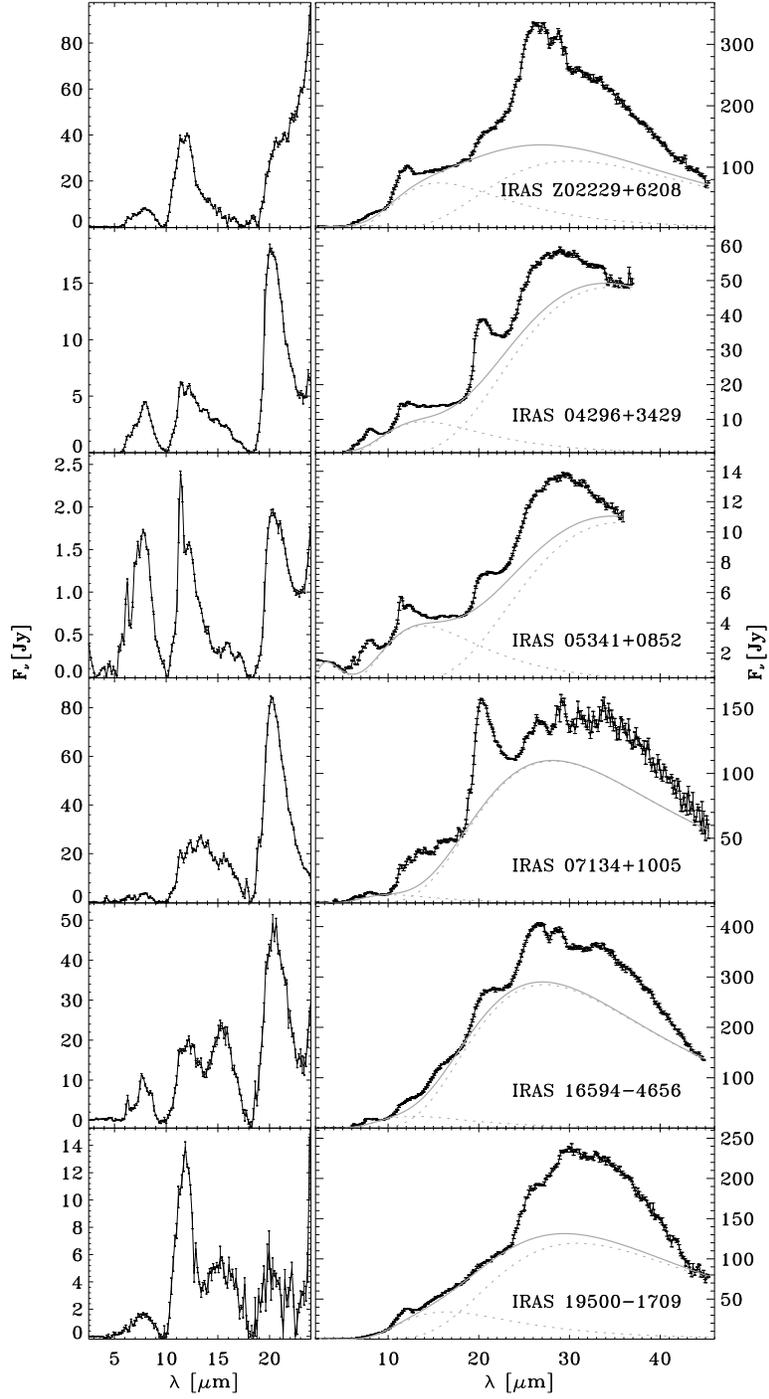}
\caption{Infrared spectra of the 21-$\mu$m sources known. A gray-body continuum is shown as a gray solid line and the single gray bodies as dashed gray lines in the right-side plots. On the left side,
 a close up of the 3--24 $\mu$m range after continuum subtraction is shown. Gray-body temperatures are listed in 
Table~\ref{tab:greytemp}.}
\label{fig:old1}
\end{figure*}

\begin{figure*}
\epsscale{0.65}
\plotone{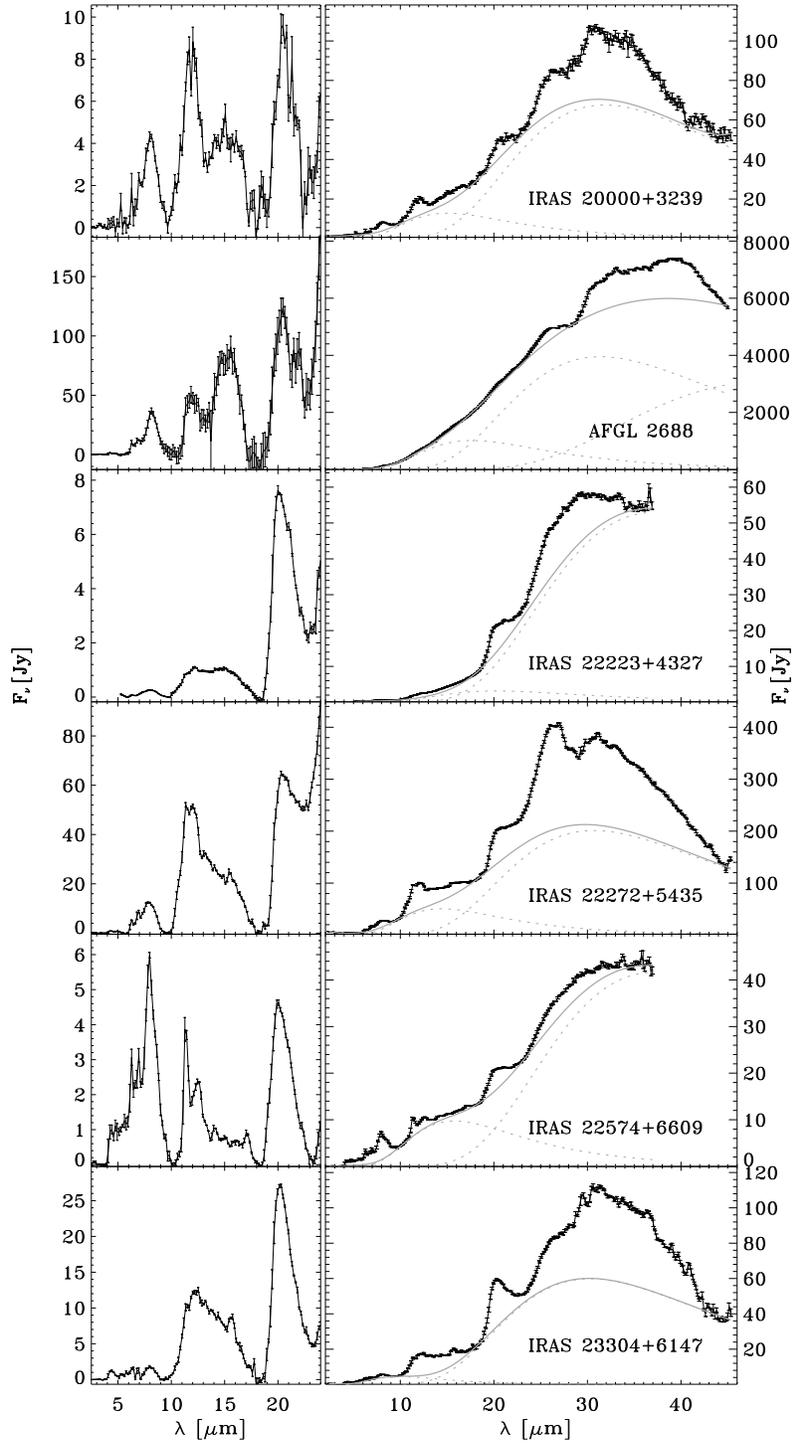}
\caption{As in Fig~\ref{fig:old1}.}
\label{fig:old2}
\end{figure*}

We have retrieved from the ISO and Spitzer archives all of the available IR spectra approximately covering the 5--40 $\mu$m 
range for the so-far known 21-$\mu$m PPNe. Table~\ref{tab:available} lists the data found in the archives and used in the following analysis.
Data for IRAS~05113+1347, 06530-0213, and 07430+1115 were not used because they do not cover the whole range from 5 $\mu$m, while
the data set for IRAS~19477+2401 was not used because of its low quality \citep{hrivnak2000}. 
When Spitzer data were available, we carried out our analysis on these, after combining them with the ISO data below 5 $\mu$m.
Post-BCD data were retrieved from the Spitzer archive, while the highly-processed data by \citet{frieswijk} were retrieved from the 
ISO web archive. Clear outliers in the ISO data were manually flagged out in SMART, then a standard $\sigma$-clip run was performed 
and the whole procedure (manual flagging and $\sigma$-clipping) was repeated a second time, before averaging the data together. 
For all of the spectra the same spectral step of 0.169~$\mu$m was set in SMART. 
The spectra thus obtained are shown in Figures~\ref{fig:old1}~and~\ref{fig:old2}.

\begin{table}[ht]
\centering 
\begin{tabular}{lcc}
\tableline\hline\noalign{\smallskip}
Target & Data sets & Range ($\mu$m) \\
\tableline\noalign{\smallskip}
Z02229+6208 & SWS\tablenotemark{a} & 2--45 \\
04296+3429 & SL\tablenotemark{b}, SH\tablenotemark{c}, LH\tablenotemark{d} & 5--37 \\
\textit{05113+1347} & SH, LH & 10--37 \\
05341+0852 & SWS, SH, LH & 2--45  \\
\textit{06530-0213} & SH, LH & 10--37  \\
07134+1005 & SWS & 2--45  \\
\textit{07430+1115} & SH, LH & 10--37  \\
16594-4656 & SWS & 2--45  \\
\textit{19477+2401} & SWS & 2--45 \\
19500-1709 & SWS & 2--45  \\
20000+3239 & SWS & 2--45  \\
AFGL 2688 & SWS & 2--45  \\
22223+4327 & SL, SH, LH & 5--37 \\
22272+5435 & SWS & 2--45  \\
22574+6609 & SWS, SH, LH & 2--45  \\
23304+6147 & SWS, SH, LH & 2--45  \\
\tableline 
\end{tabular}
\tablenotetext{a}{~Short Wavelength Spectrometer (ISO)}
\tablenotetext{b}{~Short Low IRS}
\tablenotetext{c}{~Short High IRS}
\tablenotetext{d}{~Long High IRS}
\caption{Summary of the archive observations available for all of the known 21-$\mu$m sources besides the new detections reported in this work.
Targets not included in our analysis are in \textit{italics}.}
\label{tab:available}
\end{table}

To enhance the dust features, we fitted the underlying continuum as a linear combination of four gray bodies with emissivity index $\beta=2$. In doing so, we used the IDL routine MPFITFUN \citep{mpfit}, which fits a user-defined function to the data. Although four gray bodies were always included in the fitting procedure, in most cases the best-fitting combination had a gray body multiplied by such a low factor that made it negligible, thus obtaining a best-fit curve with three gray bodies only. The same behavior was observed in trials with more than four gray-body components: only three or four of them were significant. 
As mentioned previously, the infrared continuum in post-AGB stars is due to a combination of stellar radiation and thermal emission from dust, with the latter typically much stronger than the former. To avoid physically unreasonable solutions in the fit procedure, we have therefore limited the range of values investigated by the fitting routine to $0 < T < 15000$~K, which takes into account temperature values appropriate for both the dust (dust grains sublimate around 1000--2000 K) and the central star. The temperatures
thus obtained are listed in Table~\ref{tab:greytemp} and the fitting curves are shown in Figures~\ref{fig:old1}~and~\ref{fig:old2}. We find that in all sources the emission is dominated by gray bodies reproducing dust emission (because of their temperatures) and stellar contributions are negligible.
\begin{table}\centering
 \begin{tabular}{lcccc}
\hline\hline\noalign{\smallskip}
Target & \multicolumn{4}{c}{T$_{GB}$} \\
       & \multicolumn{4}{c}{K} \\
\hline\noalign{\smallskip}
Z02229+6208 & 770     &     280    &     131    &     68  \\
05341+0852  & 639    &     148   &     57   &     --  \\
07134+1005  & 1151    &     198    &      73    &     -- \\
16594-4656  & 710    &     159    &     75    &     --  \\
19500-1709  & 1495     &     537    &      125    &     67  \\
20000+3239  & 848     &     321    &     138    &     64   \\
22272+5435  & 970    &     293    &      142    &      67  \\
22574+6609  &  350    &     130     &      55    &     --   \\
23304+6147  & 886   &     208    &     68    &     --   \\
AFGL~2688   & 233     &     117    &     65     &     40   \\
04296+3429  & 387    &     147    &     60     & -- \\
22223+4327  & 251     &     101    &     55     &  -- \\
13245-5036  &  509    &     114     &     67     & --   \\
14429-4539  &  355    &     147     &     60     & -- \\
15482-5741  &  498    &     208    &     154     & 60 \\
\hline
\end{tabular}
\caption{Temperatures of the gray bodies used to fit the underlying continuum.}\label{tab:greytemp}
\end{table}
The continuum obtained as a combination of gray bodies is typically unable to properly isolate the feature at 21~$\mu$m, 
because of the substantial contributions from the features in the 26--33 $\mu$m range, which are often blended together.

\begin{figure}
\centering
\includegraphics[width=0.45\textwidth]{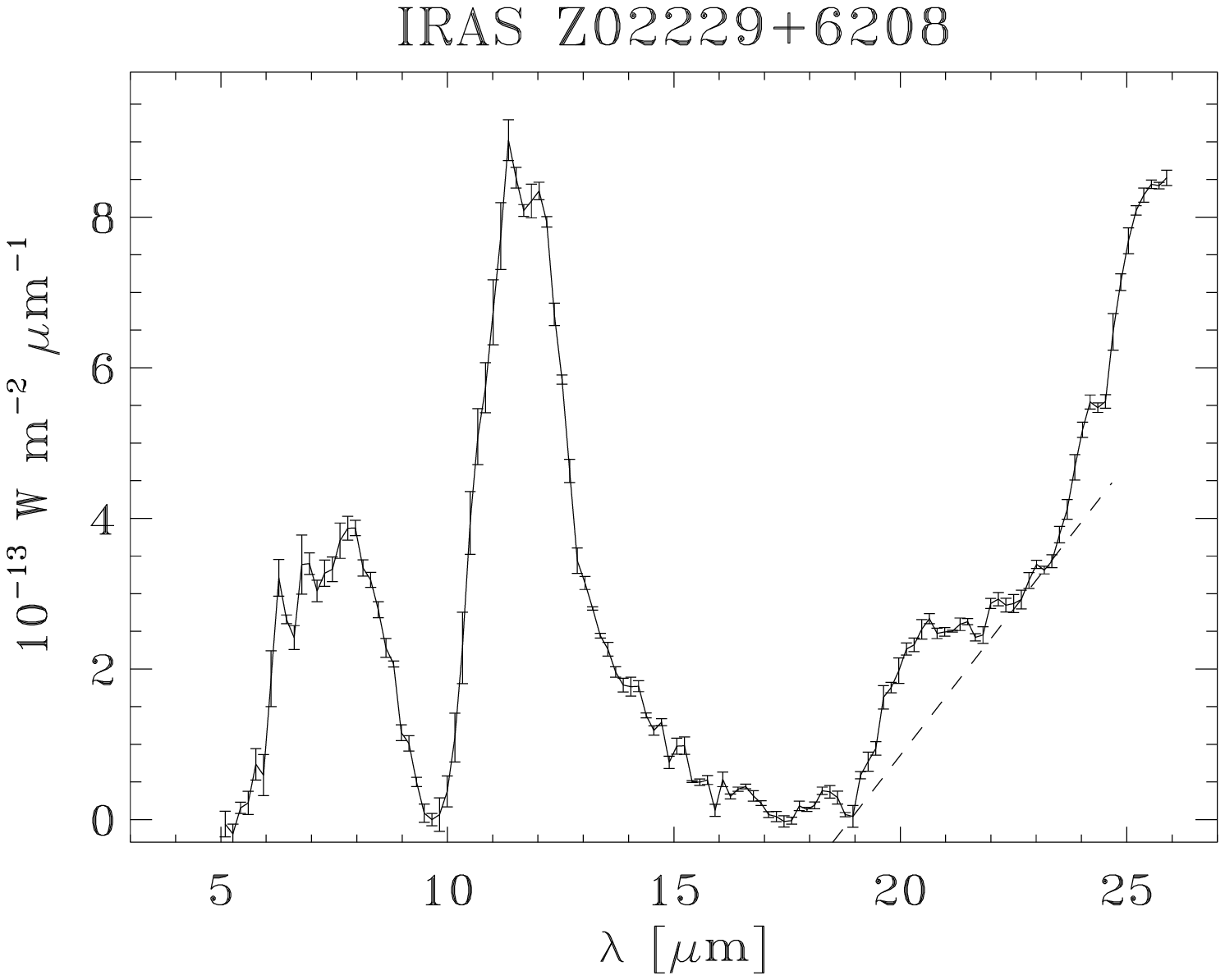}
\includegraphics[width=0.45\textwidth]{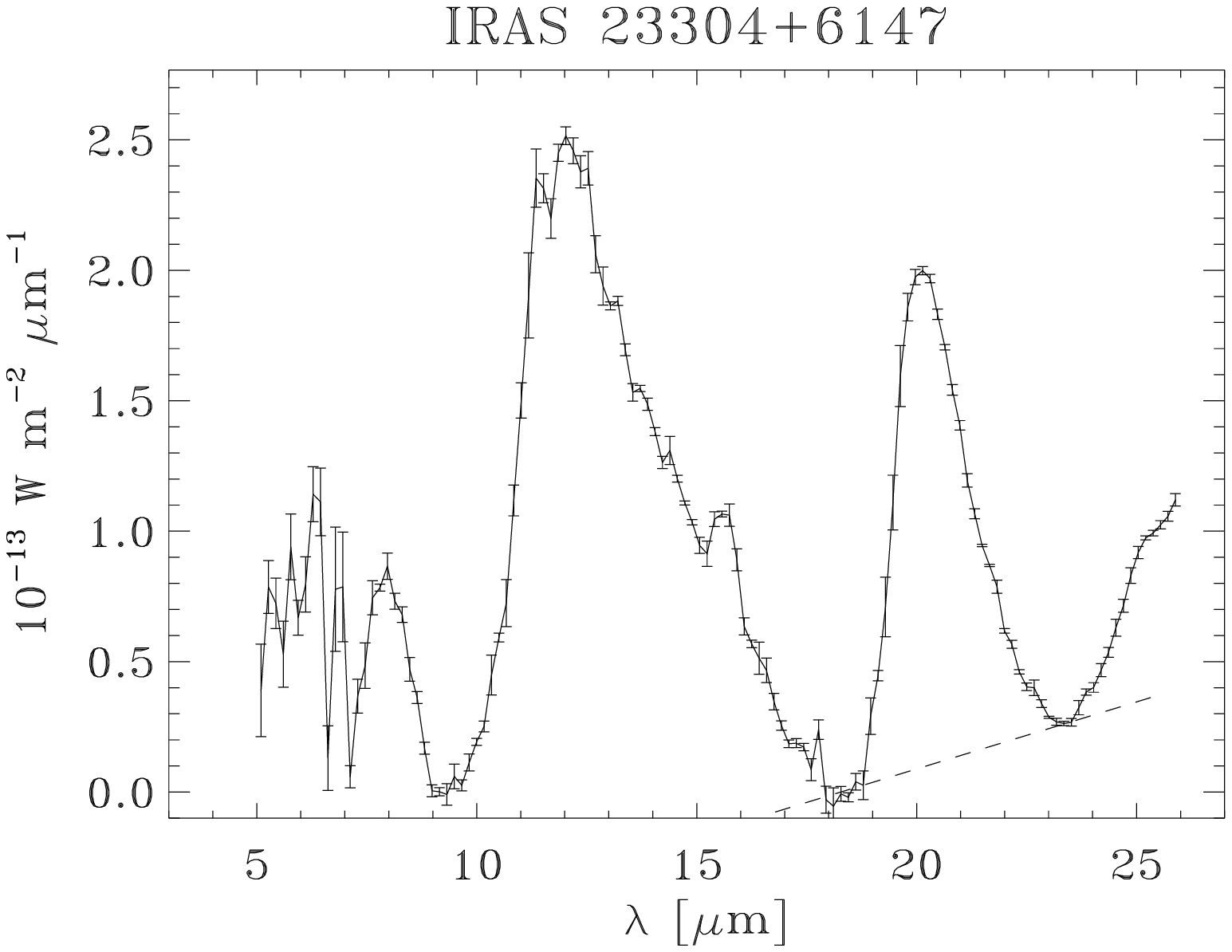}
\caption{Two examples of the straight baseline (dashed line) subtracted from the 21-$\mu$m feature.}
\label{fig:baseline}
\end{figure}

\begin{figure}
\epsscale{0.6}
\plotone{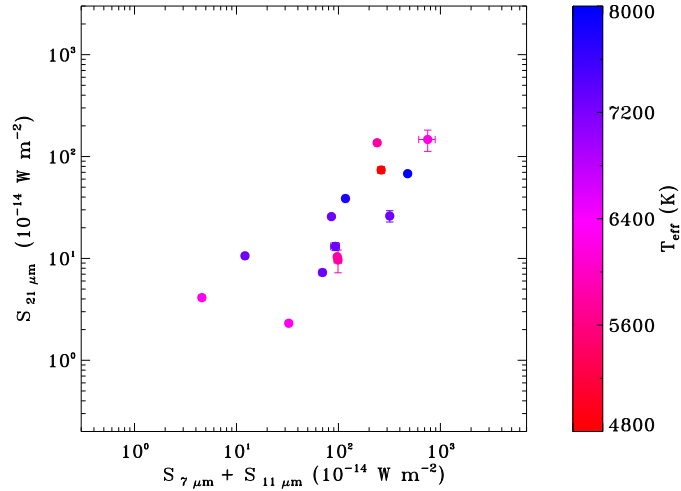}
\caption{Integrated flux emitted between 5 and 18 $\mu$m (containing the 7 and 11 $\mu$m PAH features), plotted versus the integrated flux in the 21-$\mu$m feature. The color coding depends on the temperature of the central star, with red being colder and blue hotter.}
\label{fig:intensity}
\end{figure}

Emission in the 5.5--18~$\mu$m range is present in all of the targets, as shown in the \textit{left-side} close-ups in  
Figures~\ref{fig:old1}~and~\ref{fig:old2}. After converting the spectra into W~m$^{-2}$~$\mu$m$^{-1}$, we integrated the continuum-subtracted spectra to estimate the flux in the 5.5--18~$\mu$m 
and 18.5--23.5~$\mu$m ranges (the exact edges of the ranges were  adapted to the emission in each source, turning out in differences
 of about 0.3~$\mu$m). To assess the emission in the 21-$\mu$m feature, a straight baseline was fitted to its red and blue edges and then
subtracted before performing the integration, as shown in two examples in Figure~\ref{fig:baseline}. The emission in the 5--18~$\mu$m range is mostly due to PAHs but contributions from HACs are possible, 
while the 18.5--23.5~$\mu$m range is obviously dominated by the 21-$\mu$m feature.  In Figure~\ref{fig:intensity}, 
we plot the flux estimated in the 5.5--18~$\mu$m bands, indicated as the sum of the PAH 7 and 11~$\mu$m bands, versus that in 
the 21-$\mu$m feature. 

As can be seen in the plot, a correlation between the fluxes in the two wavelength ranges is evident; 
we can calculate a linear correlation coefficient of 0.79$\pm$0.07. 
An explanation for the correlation may be that the abundances of the hydrocarbons and of the carrier of the 21-$\mu$m 
feature are linked together, possibly because of a common formation process and therefore the correlation would just reflect the different
 amounts of hydrocarbons and 21-$\mu$m carriers produced in the CSEs.

The colors of the dots in  Figure~\ref{fig:intensity} are proportional 
to the temperatures of the central stars, as derived from their spectral classifications, and range from red (coldest) to blue (hottest).
 No clear correlation is observed between the intensity of the features and the stellar temperature.

In principle, if the observed correlation were due only to a common excitation process for the molecules involved and 
then essentially to the hardening of the radiation field, we would expect to find a correlation with the temperature of the central star. 
There may be several reasons for this lack of correlation. For example, extinction effects  can play a role. Also, one of the explanations for 
the existence of the 21-$\mu$m feature implies radiation-induced decomposition of large HAC grains into smaller PAHs \citep{scott}. 
In this scenario, the carrier of the 21-$\mu$m feature would be a short-lived decomposition product.
The efficiency of the processing of large grains would depend on both T$_\mathrm{eff}$ and the location of the dust layer. A massive star 
will be hot enough to cause processing earlier than a low-mass star, with its dust still close to the central star. 
A low-mass star will reach the same temperature at a later age, with its dust farther away and will therefore need an even larger temperature
to attain comparable processing. 

\section{Summary}
One of the main results of the IRAS mission was its discovery of the existence of a broad mid-IR feature around 20~$\mu$m. The feature has been detected
almost exclusively in the circumstellar envelopes of C-rich proto-PNe, with a few weak detections in their precursors (AGB stars) 
and successors (PNe). The nature of the feature is still unclear today, but it must be linked to the C-rich nature of the sources: its carriers
are likely to be produced towards the very end of the AGB and are easily destroyed by UV radiation.

In this paper we report about the detection of three new C-rich proto-PNe that exhibit this unidentified feature through observations performed with 
the IRS instrument on-board Spitzer. Besides the 21~$\mu$m feature, our targets also show mid-IR bands from PAHs and a broad feature around 30
$\mu$m commonly attributed to MgS. As the dispersion of the dust formed during the AGB creates a very extended shell around the central star, it can be expected that evidence for such large dust regions can be found in infrared images, but our IRAC observations do not point to any extended emission. 

By a combination of catalog data and our new observations, we reconstruct the SEDs of our targets and model them with the code DUSTY.
With a r$^{-2}$ density distribution, a good match of the observational points is possible, if the model assumes
a circumstellar environment where the  mass loss process has not been constant with time. The use of FeO as carrier of the 21-$\mu$m feature
does not accurately reproduce the spectral shape of the feature. 

We have retrieved the infrared spectra of the other 21-$\mu$m sources available in the ISO and Spitzer archives. 
A gray-body continuum has been subtracted from the spectra to enhance the features. After continuum subtraction, we have calculated the flux 
in the features from 5 to 18 $\mu$m, due to hydrocarbons, and that in the 21-$\mu$m feature, finding a clear correlation between them.
We have not found any correlation with stellar temperatures. 
The correlation found between the 5--18 and 21-$\mu$m fluxes can point to a common origin
of the carriers of the features.

\acknowledgments
We are grateful to Pedro Garc{\'{\i}}a-Lario for carefully reading this manuscript and providing us with his helpful comments.\\
This work is based on observations made with the Spitzer Space Telescope, which is operated by the Jet Propulsion Laboratory, California Institute of Technology under a contract with NASA. Support for this work was provided by NASA through an award issued by JPL/Caltech.
 The Guide Star Catalogue-II is a joint project of the Space Telescope
    Science Institute and the Osservatorio Astronomico di Torino. Space
    Telescope Science Institute is operated by the Association of
    Universities for Research in Astronomy, for the National Aeronautics
    and Space Administration under contract NAS5-26555. The participation
    of the Osservatorio Astronomico di Torino is supported by the Italian
    Council for Research in Astronomy. Additional support is provided by
    European Southern Observatory, Space Telescope European Coordinating
    Facility, the International GEMINI project and the European Space
    Agency Astrophysics Division.
The DENIS project has been partly funded by the SCIENCE and the HCM plans of
the European Commission under grants CT920791 and CT940627.
It is supported by INSU, MEN and CNRS in France, by the State of Baden-W\"urttemberg 
in Germany, by DGICYT in Spain, by CNR in Italy, by FFwFBWF in Austria, by FAPESP in Brazil,
by OTKA grants F-4239 and F-013990 in Hungary, and by the ESO C\&EE grant A-04-046.
Jean Claude Renault from IAP was the Project manager.  Observations were  
carried out thanks to the contribution of numerous students and young 
scientists from all involved institutes, under the supervision of  P. Fouqu\'e,  
survey astronomer resident in Chile.  
This publication makes use of data products from the Two Micron All Sky Survey, which is a joint project of the University of Massachusetts and the Infrared Processing and Analysis Center/California Institute of Technology, funded by the National Aeronautics and Space Administration and the National Science Foundation.
This research is based in part on observations with AKARI, a JAXA project with the participation of ESA.\\



{\it Facilities:} \facility{ISO (SWS)}, \facility{Spitzer}.

\clearpage

\clearpage






\end{document}